\documentclass[sigconf]{acmart}

\usepackage{latexsym}
\usepackage{makecell}
\usepackage[english]{babel}

\AtBeginDocument{%
  \providecommand\BibTeX{{%
    \normalfont B\kern-0.5em{\scshape i\kern-0.25em b}\kern-0.8em\TeX}}}

\copyrightyear{2021}
\acmYear{2021}
\setcopyright{acmcopyright}\acmConference[CIKM '21]{Proceedings of the 30th ACM International Conference on Information and Knowledge Management}{November 1--5, 2021}{Virtual Event, QLD, Australia}
\acmBooktitle{Proceedings of the 30th ACM International Conference on Information and Knowledge Management (CIKM '21), November 1--5, 2021, Virtual Event, QLD, Australia}
\acmPrice{15.00}
\acmDOI{10.1145/3459637.3482097}
\acmISBN{978-1-4503-8446-9/21/11}

\newcommand{\figref}[1]{Figure \ref{#1}}

\newcommand{\tabref}[1]{Table \ref{#1}}

\newcommand{\cut}[1]{}

\begin{document}
\fancyhead{}

\title{Enriching Ontology with Temporal Commonsense \\ for Low-Resource Audio Tagging}


\author{Zhiling Zhang}
\affiliation{%
  \institution{Shanghai Jiao Tong University}}
\email{blmoistawinde@sjtu.edu.cn}
\orcid{0000-0002-8081-704X}

\author{Zelin Zhou}
\affiliation{%
  \institution{Shanghai Jiao Tong University}}
\email{ze-lin@sjtu.edu.cn}

\author{Haifeng Tang}
\affiliation{%
  \institution{CMB Credit Card Center}}
\email{thfeng@cmbchina.com}

\author{Guangwei Li*}
\affiliation{%
  \institution{Shanghai Jiao Tong University}}
\email{ligw20@sjtu.edu.cn}

\author{Mengyue Wu* $\dagger$}
\thanks{* authors are with MoE Key Lab of Artificial Intelligence and X-LANCE Lab}
\affiliation{%
  \institution{Shanghai Jiao Tong University}}
\email{mengyuewu@sjtu.edu.cn}
\thanks{$\dagger$ Corresponding author, supported by National Natural Science Foundation of China (No.61901265), 
Shanghai Pujiang Program (No.19PJ1406300), and Shanghai Municipal Science and Technology Major Project (2021SHZDZX0102).}

\author{Kenny Q. Zhu $\ddagger$}
\affiliation{%
  \institution{Shanghai Jiao Tong University}}
\email{kzhu@cs.sjtu.edu.cn}
\thanks{$\ddagger$ Corresponding author, and is partially supported by
SJTU-CMB Credit Card Center Joint Research Grant and SJTU Medical-Engineering
Cross Disciplinary Scheme.}

\begin{abstract}
  Audio tagging aims at predicting sound events occurred in a recording. Traditional models require enormous laborious annotations, otherwise performance degeneration will be the norm. Therefore, we investigate robust audio tagging models in low-resource scenarios with the enhancement of knowledge graphs. Besides existing ontological knowledge, we further propose a semi-automatic approach that can construct temporal knowledge graphs on diverse domain-specific label sets. Moreover, we leverage a variant of relation-aware graph neural network, D-GCN, to combine the strength of the two knowledge types. Experiments on AudioSet and SONYC urban sound tagging datasets suggest the effectiveness of the introduced temporal knowledge, and the advantage of the combined KGs with D-GCN over single knowledge source.
\end{abstract}



\maketitle

\section{Introduction}
\label{sec:intro}

Audio tagging is the task to label the sound recordings with representative tags like sound events. It can be used in many applications like music tagging \citep{fu2010survey}, sound retrieval \citep{font2018sound} and urban sound planning \citep{bello2019sonyc}. However, the manual labeling required by a well-performing tagging model can be heavy and expensive. Compared to annotating for images classification, annotators have to spend substantially more time on finishing the whole recording before tagging. Consequently, even the largest dataset for audio tagging, AudioSet \citep{gemmeke2017audio}, is orders of magnitude smaller than image classification datasets \citep{deng2009imagenet, thomee2016yfcc100m}, and many other common datasets are even smaller, like SONYC \citep{bello2019sonyc} with only 2,351 training recordings. Therefore, when we want to tailor a new dataset for particular needs, the data size tends to be limited, and thus robust methods in such low-resource scenarios are largely desired.

Structured knowledge has been shown effective in the low-resource setting of many tasks \citep{shwartz2020unsupervised, feng2020scalable, ji2020language, li2013data, zhang2013automatic}, while the exploration of knowledge is still limited for audio tagging. These works mainly focus on ontological knowledge \citep{jati2019hierarchy, sun2020ontology, shrivaslava2020mt}. For example, AudioSet involves \textit{IsA} relations between sound events, such as \textit{Speech IsA Human voice}. Models enhanced with such knowledge can effectively capture the similarity between tags of similar category. However, ontological knowledge failed to cover other relations between different categories, like the frequent \textit{Co-occurrence} of \textit{Car} sounds and \textit{Engine} sounds. Such relationships are still under-explored.

\begin{figure}[htbp]
  \setlength{\abovecaptionskip}{0.cm}
\setlength{\belowcaptionskip}{-0.5cm}
  \centering
  \includegraphics[width=\linewidth]{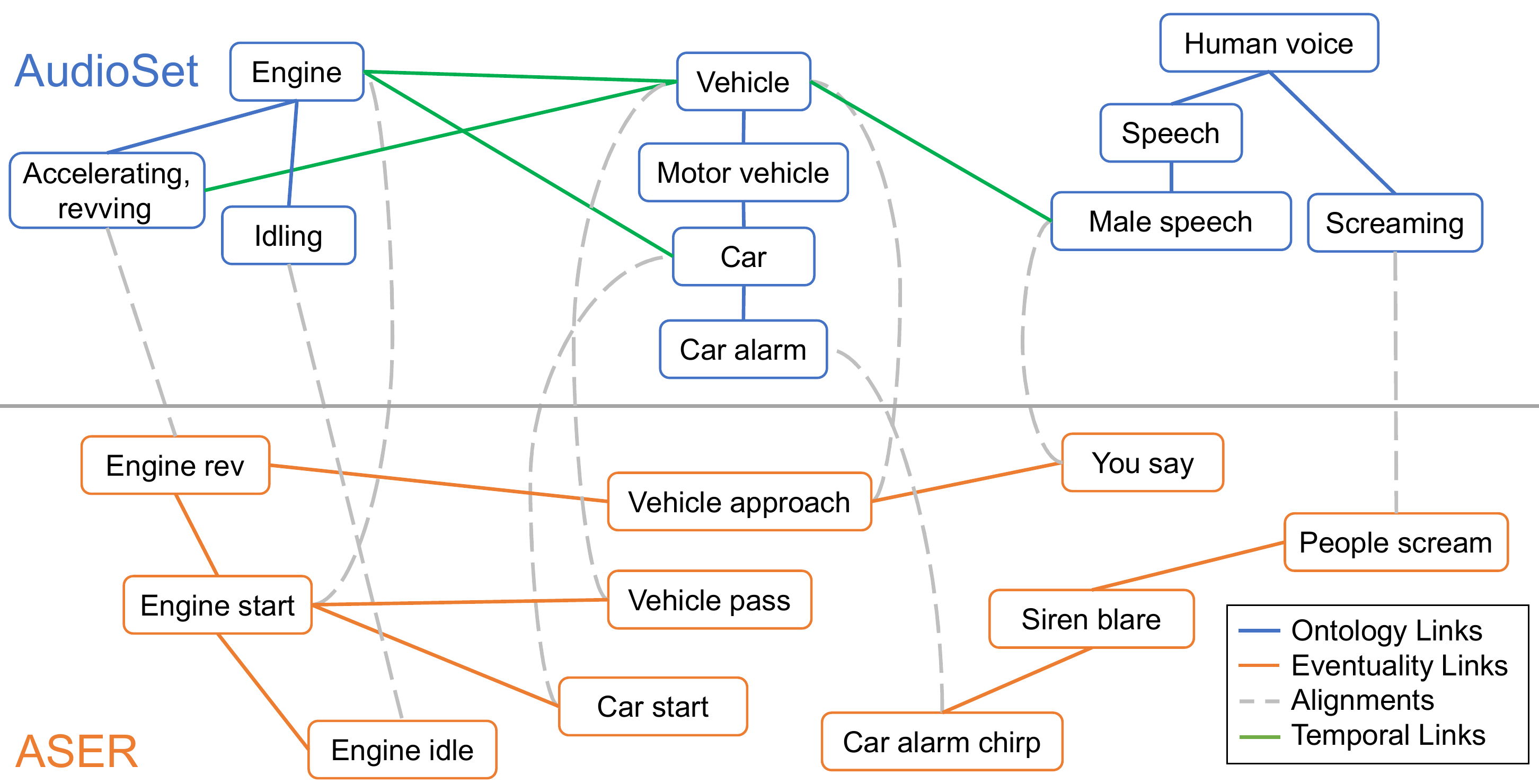}
  \caption{Aligning ASER to AudioSet. We establish an enriched link between two AudioSet events if their corresponding ASER events have temporal links.}
  \label{fig:align}
\end{figure}

To extend the dimensions of available knowledge for audio analysis, 
we propose to construct temporal knowledge graph on top of pre-defined audio tags 
with the knowledge transferred from ASER \citep{zhang2020aser}, 
a large-scale commonsense knowledge graph about the relations between events extracted 
from textual corpus. Here we pay special attention to temporal relations 
like \textit{Precedence} and \textit{Conjunction}, 
which we will collectively refer to as ``temporal links''.
They can provide hints whether 2 events might co-occur in a short period, 
as is the case for multi-label audio tagging. 
This kind of knowledge is similar to the sound event co-occurrence graph mined 
from audio dataset \citep{shrivaslava2020mt, wang2020modeling} to some extent, 
but they come from different sources, i.e., text vs. audio. 
Algorithms to mine sound event co-occurrence from audio data
rely on dataset-specific hyperparameter tuning, 
and large-scale, accurate, multi-label annotations, 
which can be hard to obtain due to missing labels problems~\citep{fonseca2020addressing},
or from single-label or low-resource datasets.

The knowledge construction requires the alignments between ASER events and domain-specific 
label sets, usually accompanied with ontologies, such as AudioSet (\figref{fig:align}).
We will then add an enriched link between two Audioset events if any pair of their aligned events in ASER have temporal links.
There are many challenges in establishing the alignments. 
First, there are mismatches between the event representations. For example, 
events are usually named with representative noun phrases or verb phrases in AudioSet, 
while ASER provides a finer representation of short clauses. 
Therefore, one AudioSet event can be aligned to multiple related ASER events, 
like \textit{Vehicle} to \textit{Vehicle approach} and \textit{Vehicle pass}. 
Moreover, the alignments should be decided according to acoustic relatedness rather 
than lexical or semantical similarity alone. For instance, we should not 
treat \textit{I see engine} as related to the sound event of \textit{Engine} 
despite the shared object, as the former event doesn't make a sound. In contrast, 
we should link \textit{Male Speech} to \textit{You say} although they have no words
in common.

In this work, we propose a semi-automatic approach to align the events in ASER and those in
the target label set. We use heuristics to automatically extract audible events in
ASER that are synonymous to those events in AudioSet, with the help of Word Sense Disambiguation (WSD) \citep{lesk1986automatic} and WordNet \citep{miller1995wordnet}. For AudioSet, candidate alignments are also verified
manually to ensure the quality of final temporal KG.
The resulting temporal KGs can be directly incorporated into existing Graph Convolutional Network (GCN)-based models. We conduct experiments on two audio tagging datasets in different domains: the open domain AudioSet \citep{gemmeke2017audio} and SONYC urban sound tagging dataset \citep{bello2019sonyc}. Results suggest the usefulness of temporal knowledge in low resource settings. Nonetheless, simple combination of temporal and ontological knowledge sources in one graph doesn't provide performance gain. We hypothesize that a single graph can't model the heterogeneity of the two knowledge types properly. Inspired by R-GCN \citep{schlichtkrull2018modeling}, we further propose D-GCN that applies relation-specific transformation for two knowledge types.

In summary, our contributions are: 
(1) We propose a semi-automatic approach to construct temporal commonsense knowledge graph to enrich different audio tagging ontologies like AudioSet and SONYC.
(2) We propose to use a variant of R-GCN of two relation types, D-GCN, 
to leverage the heterogeneity of ontological and temporal knowledge.
(3) Experiments on AudioSet and SONYC dataset in low-resource settings show 
that temporal KG-enhanced model outperforms backbone model without KG, and 
combining ontology and temporal KG with D-GCN can provide further improvement. \footnote{Code and data: \url{https://github.com/blmoistawinde/dgcn_tagging}}

\section{Method}
\label{sec:method}

\subsection{Knowledge Resources}
\label{sec:resources}

\textit{AudioSet} \citep{gemmeke2017audio} is a hierarchically structured ontologies comprised of 632 audio events. It is by far the largest ontology, covering most common sound events. 
\textit{SONYC} \citep{bello2019sonyc} is a two-level taxonomy consists of 8 coarse level tags, 23 fine level tags about urban sounds. This taxonomy is smaller due to its specific target of detecting urban noises.
\textit{ASER} \citep{zhang2020aser} is a large-scale eventuality knowledge graph extracted from textual corpus. Each eventuality is represented as a short clause containing lemmatized words of subject, verb, object, etc. In this work, we will use its core version with 27,565,673 event nodes and 8,834,257 relation edges of 15 types. We will focus on temporal relations types like \textit{Precedence} and \textit{Conjunction}.

\subsection{Knowledge Construction}
\label{sec:enrich}
\begin{figure}[htbp]
\setlength{\abovecaptionskip}{0.cm}
\setlength{\belowcaptionskip}{-0.5cm}
  \centering
  \includegraphics[width=1\linewidth]{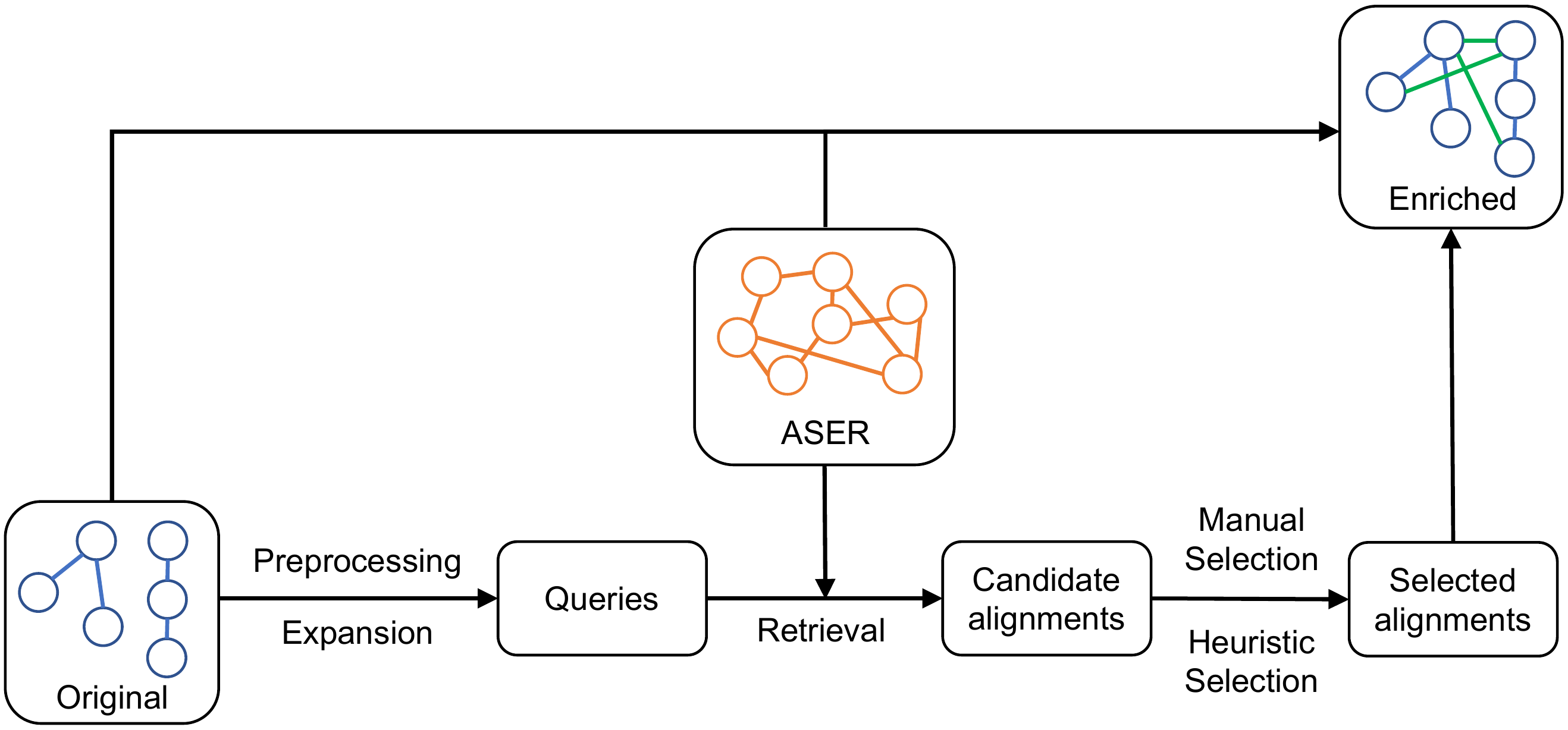}
  \caption{Illustration of alignment pipeline.}
  \label{fig:pipeline}
  \end{figure}

In this section, we will introduce how to perform the alignment between an audio event label set and ASER, and construct the temporal KG to enrich the existing ontologies (Figure \ref{fig:pipeline}) . Without loss of generality, we will mainly describe the procedure with AudioSet as target, and explain the difference for SONYC when necessary. 

\subsubsection{Preprocessing and Expansion}

First, we conduct preprocessing to deal with the morphological differences between two type of representations. The words in ASER events are already lemmatized, while the verbs in AudioSet tags are not. Therefore, we enforce lemmatization on AudioSet tags, and also do lowercasing, remove the parentheses and stopwords. 

After that, since there are lexical variations in the expressions of similar audio events in ASER, we propose to expand a single AudioSet tag into multiple equivalent queries to improve the recall of alignment candidates retrieval (\S \ref{sec:cand-retrieval}). 

First, some of the tags already contain parallel concepts or synonyms, like ``Roaring cats (lions, tigers)'', and we will split it into multiple queries with one concept each, like ``roaring cats'', ``lions'' and ``tigers''. 
Moreover, for those tags without provided synonyms, we can also extract their synonyms from the lexical database WordNet \citep{miller1995wordnet}. To leverage the knowledge from WordNet, we applied the Lesk algorithm \citep{lesk1986automatic} for Word Sense Disambiguation (WSD) to link each word to its corresponding Wordnet synset. 
There are also ambiguous events in AudioSet with multiple father events. For example, the event ``Hiss'' is a child event of ``Cat'', ``Snake'' and ``Steam'', and the acoustic property of the event might be different under different father events. Hence, we produce different queries by pairing such event with each of its father events, so the expanded queries for the above example would be ``cat hiss'', ``snake hiss'' and ``steam hiss''.

\subsubsection{Alignment Candidates Retrieval}
\label{sec:cand-retrieval}

Since ASER contains numerous events, we need to retrieve a small number of likely candidates before alignment selection. At first, we filter out some unlikely candidates by excluding events with noisy patterns like duplicate verbs (``I say say''), and infrequent ones (with frequency < 5). We then retrieve the top 10 events for each query with ElasticSearch. First, we adopt pure text-matching, and give higher weight to the matching of verbs, as verbs are the key components of events. Moreover, since we noticed that the above text matching approach may sometimes retrieve infrequent events that is either too rare or too specific, we also supplemented the above method with another weighting scheme that adds additional weights according to event frequency. Consequently, retrieval results will contain more general and frequent events, which will have more linked relations to leverage. The results from both weighting schemes are combined to balance accuracy and frequency. 
Finally, the retrieved ASER events with the multiple queries of the same AudioSet event are
aggregated as the candidates for alignment. As a result, each event has 31.3 candidates in average with a minimum of 2 and maximum of 190. 

\subsubsection{Selection} 

Now we need to select events that are precisely related to a sound event in terms of acoustic property from the retrieved candidates. 

For the alignments to AudioSet, we manually annotate them ourselves to ensure the reliability. We need to decide for each candidate eventuality of an AudioSet event, whether they are related, unrelated or that the relation is ambiguous. The event name, description, father/child events, corresponding queries and example videos containing that event are shown to aid the decision. Among all the annotated alignments, about 31.96\% are considered related, 13.52\% ambiguous and 54.51\% unrelated. To ensure the precision of the results, we only use the ``related'' alignments later.

For the alignments to SONYC, we completely automate the selection process. We observe that
most candidates for certain specific event labels (like ``Reverse Beeper'') are already of acceptable quality. Therefore, we preserve all the candidates of specific labels except ``other/unknown'' labels and under-specified labels like ``machinery impact''. 

\subsubsection{Construction}

The relations between ASER events are transferred to their corresponding AudioSet events through the alignments, and their relations will be aggregated. For example, let's assume that an AudioSet event $a_1$ is aligned to events $e_{11}, e_{12}$, and another event $a_2$ is aligned to $e_{21}$. In ASER, $e_{11}$-$e_{21}$ has the relation \textit{(`Co\_Occurrence', `Conjunction')}, and $e_{12}$-$e_{21}$ has the relation \textit{(`Co\_Occurrence', `Precedence')}. Then $a_1$-$a_2$ will have the aggregated relation \textit{(`Co\_Occurrence', `Conjunction', `Precedence')}. The resulting KG will inherit various types of relations from ASER, and we will only use \textit{Conjunction} and \textit{Precedence} in later experiments as other relations are either not temporal or too sparse.
\subsection{Double GCN (D-GCN)}

Typically, CNN-based models \citep{kong2020panns} are used for audio tagging. To leverage KGs, we roughly follow \citep{wang2020modeling} to add a GCN component to learn the representation of labels. Then the audio representations from the CNN encoder are dot-producted with the learned label representations, followed by a sigmoid layer to get the prediction. 

Either the ontology or temporal KG alone can be incorporated into GCN-based models for audio tagging. However, a single GCN may not be suitable to handle the heterogeneity of two knowledge types when we want to combine their strength in one model. We thus draw inspiration from R-GCN \citep{schlichtkrull2018modeling}, which introduced relation-specific transformations.

\begin{equation}
\setlength{\abovedisplayskip}{3pt}
\setlength{\belowdisplayskip}{3pt}
  h_i^{l+1} = \sigma(\sum_{r \in \mathcal{R}} \sum_{j \in N^r_i} \frac{1}{c_{ij, r}} W^{l}_r h^{l}_j)
\end{equation}
where $h_i^{l}$ are the features of node $i$ at the $l$-th layer, $\mathcal{R}$ is the set of relations, $N^r_i$ is the indices of the neighbors of $i$ via edges with relation $r$ (self-loop included), $c_{ij, r}$ is a normalization constant, and $W^{l}_r$ is the specific weight matrix for relation $r$ at the $l$-th layer. The original R-GCN also introduce basis function decomposition and block decomposition for $W^{l}_r$ as a regularization to reduce parameters and prevent overfitting on rare relations.

Different from the original R-GCN, which aims to tackle the completion of knowledge bases with over 1,000 relation types, we only need to model 2 kinds of relations with relatively balanced numbers. Therefore, we don't apply the regularizations and further add relation specific bias terms, to improve the model's expressiveness on each relation. We will refer to this variant as D-GCN.

\section{Experiments}
\label{sec:experiments}


We experiment our methods on datasets with corresponding ontologies to verify the effectiveness of either ontological or temporal knowledge alone, and their combination.

\textit{AudioSet} is a large-scale multi-label audio tagging dataset collected from Youtube videos with the annotations of 527 categories out of the 632 tags defined in the ontology. Most recordings are processed into 10 seconds single-channel 16kHz, 16-bit wave format. Due to the changes of videos, it is not possible to recover the whole dataset. We downloaded 19,400 (87.5\%), 1,851,420 (90.7\%), and 17,756 (87.2\%) recordings for the balanced train, full train, and evaluation set, respectively. To simulate the low-resource scenario, we randomly sample 1\% of the unbalanced set as the training set, which has 18,514 samples, where 134 classes have no more than 5 samples, and 10 classes have no training sample. We also sample 5\% and 10\% sets for comparison. We use the commonly used mAP, mAUC as the evaluation metrics. 
\textit{SONYC} is the multi-label Urban Sound Tagging dataset used in DCASE 2019 Task5 (D19T5). It contains 2,351 train recordings and 443 validate recordings, which can be considered relatively low-resource. All recordings are 10 seconds single-channel 44.1kHz, 16-bit wave format. We use the official metrics, micro AUPRC and macro AUPRC.

\subsection{Experimental Setup}

\paragraph{Audio Features} For SONYC, we adopt a similar setting as \citep{kong2019cross}, all audios are re-sampled to 32 kHz and 64-Mel-bin log-Mel spectrograms are used to to represent the audios. The window size is 1024 samples, the hop size of 500 samples, and cut-off frequencies of 50 Hz to 14 kHz. 
For AudioSet, our setting is similar to \citep{kong2020panns}, all audios are re-sampled to 16 kHz and represented as 64-Mel-bin log-Mel spectrograms. The window size is 512 samples, the hop size of 160 samples, and cut-off frequencies of 50 Hz to 8 kHz.

\paragraph{Models and Baselines} We adopt standard CNN models as our baseline, that is, the CNN9 model use in \citep{kong2019cross} for SONYC and the CNN14 (16kHz) in \citep{kong2020panns} for AudioSet. We also introduce the SOTA method AT-GCN \citep{wang2020modeling} as baseline, which is based on co-occurrence graph mined from the whole AudioSet annotation, and uses tuned, dataset-specific hyperparameter for edge thresholding and smoothing. It is thus not directly applicable to SONYC. Our models include GCN(ASER), GCN(AudioSet), GCN(ASER+AudioSet), which refers to single GCN with temporal knowledge, AudioSet ontology, and their combination. D-GCN denotes double-GCN with 2 types of knowledge. We replace AudioSet with SONYC's ontology (OT) for experiments on SONYC dataset. We use batch size of 32 for all models and the learning rate is 1e-3 for all models except D-GCN using 3e-4.

\subsection{Results}

\subsubsection{AudioSet}

\begin{table}[tbp]
  \setlength{\belowcaptionskip}{-0.cm}
  \centering
  \small
  \begin{tabular}{lcccc}
      \hline
      {} & \multicolumn{2}{c}{Balance} & \multicolumn{2}{c}{Unbalance (100\%)} \\
      \cline{2-3}\cline{4-5} 
      Methods & mAP & mAUC & mAP & mAUC \\
      \hline
      CNN14 & 0.2441 & 0.8930 & 0.4090 & 0.9669 \\
      \hline
      AT-GCN & 0.2510 & 0.9278 & 0.4095 & 0.9664 \\
      GCN(ASER) & 0.2500 & 0.9283 & 0.3994 & 0.9660 \\
      GCN(AudioSet) & 0.2543 & \textbf{0.9420} & 0.4063 & 0.9665 \\
      GCN(ASER+AudioSet) & 0.2490 & 0.9277 & 0.3999 & \textbf{0.9690} \\
      D-GCN & \textbf{0.2554} & 0.9377 & \textbf{0.4109} & 0.9648 \\
      \hline
  \end{tabular}
  \caption{\label{tab:bal-unbal} Results on AudioSet evaluation set with models trained on balanced and unbalanced set.}
\end{table}

\begin{table}[tbp]
  \setlength{\belowcaptionskip}{-0.cm}
  \centering
  \small

  \begin{tabular}{lccc}
      \hline
      Methods &     1\% &     5\% &    10\% \\
      \hline
      CNN14                 & 0.1118 & 0.2343 & 0.2770 \\
      \hline
      AT-GCN                & 0.1243 & 0.2331 & 0.2785 \\
      GCN(ASER)             & 0.1252 & 0.2269 & 0.2735 \\
      GCN(AudioSet)         & 0.1280 & 0.2336 & 0.2747 \\
      GCN(AudioSet+ASER)    & 0.1214 & 0.2283 & 0.2741 \\
      D-GCN  & \textbf{0.1283} & \textbf{0.2387} & \textbf{0.2799} \\
      \hline
      D-GCN rel. improvement & 14.73\% & 1.88\% & 1.05\% \\
      \hline
  \end{tabular}   
  \caption{\label{tab:low-resource-map} mAP for each model trained on different portion of the unbalanced set, and the relative improvement(\%) of D-GCN over CNN14 backbone.}
  \vspace{-4mm}
\end{table}

\tabref{tab:bal-unbal} shows the performance of models trained on the official balanced and unbalanced set, while \tabref{tab:low-resource-map} shows the results on sampled subsets. We can see that all GCN-based models significantly outperform the baseline CNN14 on balanced and low-resource (1\%) set, suggesting the usefulness of the knowledge sources including the newly proposed temporal knowledge in low-resource scenarios. As the size of training data grows, the advantage of GCN models ceases to exist, expect for AT-GCN, possibly due to its knowledge of the co-occurrences on the whole training set, which matches more with the larger training data. D-GCN performs consistently better than single GCN with one KG or the simple addition of both KGs, showing the effectiveness of the separate relation modeling, and it also outperforms CNN14 by mAP in all settings despite the diminishing gain.

\begin{figure}[htbp]
\setlength{\abovecaptionskip}{0.cm}
\setlength{\belowcaptionskip}{-0.5cm}
\centering
\includegraphics[width=0.9\linewidth]{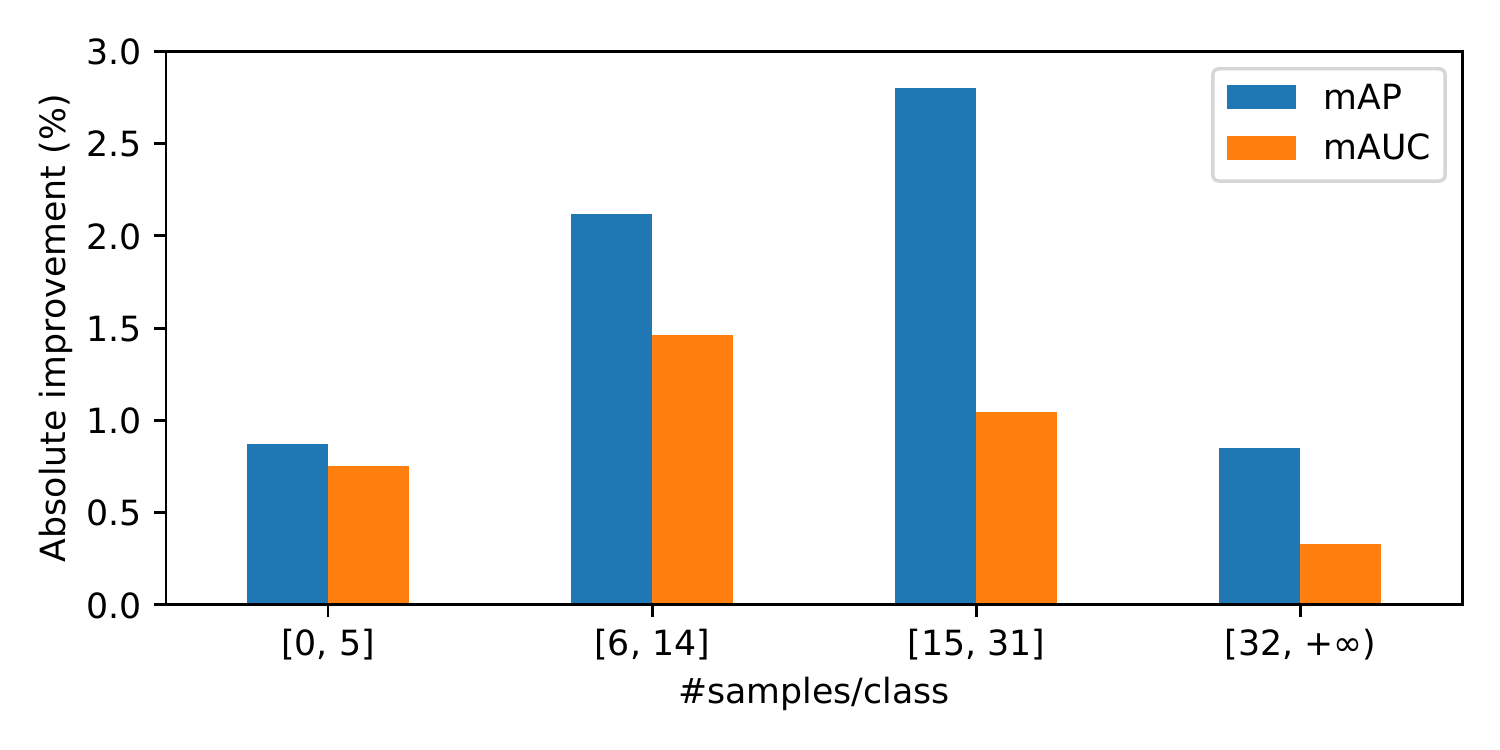}
\caption{Absolute improvement of D-GCN over CNN14 on classes with different number 
of training samples.}
\label{fig:delta_map_auc}
\end{figure}

To study the reason for the effectiveness of GCN models in low-resource scenario (1\% set) and their degeneration in large-data settings, we divide the classes into groups according to the numbers of training samples, and calculate D-GCN's improvement over baseline on these groups. From Figure \ref{fig:delta_map_auc}, we can see that D-GCN can benefit classes with extremely few samples ([0, 5]), and the gain is the highest on classes with moderate number of samples, but not on the most prevalent classes. We may conclude that the prior knowledge in KG can effectively help the model learn the dependency between labels especially for the few-shot ones. However, as we have more resources, the large backbone model may be capable of learning such relations without KG, which explains why the advantage of GCN-based models would shrink.   

\subsubsection{SONYC}

\begin{table}[tbp]
  \setlength{\belowcaptionskip}{-0.cm}
    \centering
    \small
    \begin{tabular}{lcccc}
        \hline
        {} & \multicolumn{2}{c}{Fine-level} & \multicolumn{2}{c}{Coarse-level} \\
        \cline{2-3}\cline{4-5} 
        Methods & Mi AUPRC & Ma AUPRC & Mi AUPRC & Ma AUPRC \\
        \hline
        CNN9 & 0.675 & 0.493 & 0.808 & 0.580 \\
        GCN(ASER) & 0.703 & 0.459 & 0.822 & 0.548 \\
        GCN(OT) & 0.680 & 0.494 & 0.821 & 0.596 \\
        GCN(ASER+OT) & 0.706 & 0.492 & \textbf{0.823} & 0.616 \\
        D-GCN & \textbf{0.709} & \textbf{0.516} & 0.820 & \textbf{0.647} \\
        \hline
    \end{tabular}
    \caption{\label{tab:SONYC} Results on SONYC validate set, OT: SONYC ontology, Mi: Micro, Ma: Macro.}
    \vspace{-4mm}
  \end{table}

The results on the SONYC dataset is shown in Table \ref{tab:SONYC}. Similar to AudioSet (1\%), all GCN models significantly outperform baseline by the main metric Micro AUPRC. The temporal knowledge of ASER seems to be more useful here compared to ontology, as the ontology for SONYC is more sparse, and the labels for each level are predicted separately, so that they don't co-occur. D-GCN again gives consistently best or competitive performance on both level, suggesting the generalizability of this method on effectively combining the strength of two knowledge types.

\section{Conclusion}
We investigated the efficacy of KG-enhanced models for low-resource audio tagging. We proposed a semi-automatic procedure to build temporal knowledge graph in multiple domains to enrich existing ontologies. We further proposed D-GCN to effectively combine knowledge of two distinct types. Results on AudioSet and SONYC showed that GCN-based model with the introduced temporal knowledge can significantly outperform baseline without prior knowledge especially in low-resource settings, and D-GCN model with combined knowledge can provide further improvement over models with single type of knowledge.


\bibliographystyle{ACM-Reference-Format}
\bibliography{cikm}

\appendix

\end{document}